\documentclass[aps, pra, amsmath,amssymb,footinbib, twocolumn,superscriptaddress, 10pt]{revtex4-1}

\usepackage{graphicx}
\usepackage{bm}
\usepackage{amssymb}
\usepackage{color}
\usepackage{amsmath}
\usepackage{amstext}
\usepackage{amsthm}
\usepackage{latexsym}
\usepackage{physics}
\usepackage[dvipsnames]{xcolor}
\usepackage[colorlinks=true,citecolor=RedViolet,urlcolor=purple]{hyperref}
\usepackage{bbold}
\usepackage{float}
\usepackage{tikz}
\usepackage{enumerate}
\usepackage{libertinus}
\usepackage{upgreek}
\usepackage{mathrsfs}

\definecolor{pink}{rgb}{1,0.078,0.57}

\definecolor{green}{rgb}{0,0.7,0.9}

\def\x{{\rm\bf x}}
\def\y{{\rm\bf y}}

\usetikzlibrary{positioning}

\newcommand{\mc}[1]{\mathcal{#1}}
\newcommand{\mr}[1]{\mathrm{#1}}
\newcommand{\ts}[1]{\textsc{#1}}
\newcommand{\mf}[1]{\mathfrak{#1}}
\newcommand{\ha}[1]{\hat{#1}}

\newcommand{\mbb}[1]{\mathbb{#1}}
\newcommand{\bs}[1]{\boldsymbol{#1}}

\newcommand{\ep}[1]{^{(#1)}}
\graphicspath{{../figs/}}

\newcommand{\Id}{\hat{\mathbb{1}}}

\newcommand{\msc}[1]{\mathscr{#1}}

\newcounter{am}

\makeatletter
\g@addto@macro\normalsize{%
  \setlength\abovedisplayskip{7pt plus 4pt minus 4pt}%
  \setlength\belowdisplayskip{7pt plus 4pt minus 4pt}%
  \setlength\abovedisplayshortskip{0pt plus 2pt}%
  \setlength\belowdisplayshortskip{4pt plus 2pt minus 2pt}%
  \setlength\jot{2pt}
}
\makeatother

\begin{document}
\title{On information propagation in Krylov subspaces}

\author{Pablo Martinez-Azcona \href{https://orcid.org/0000-0002-9553-2610}{\includegraphics[scale=0.04]{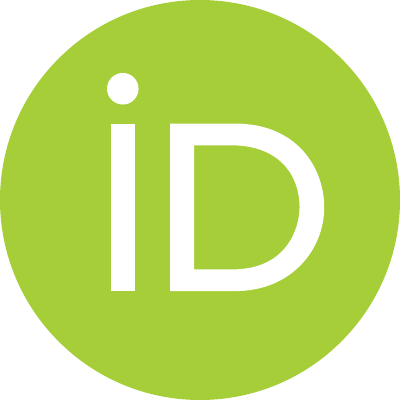}}}
\thanks{Corresponding authors:\\ pablo.martinez-azcona@tum.de, sam310@cam.ac.uk}
\affiliation{Technical University of Munich, TUM School of Natural Sciences,
Physics Department, 85748 Garching, Germany}
\affiliation{Walther-Meißner-Institut, Bayerische Akademie der Wissenschaften, 85748 Garching, Germany }
\affiliation{Munich Center for Quantum Science and Technology (MCQST), 80799 Munich, Germany}

\author{Maximilian Meyer-Mölleringhof \href{https://orcid.org/0009-0002-2237-8058}{\includegraphics[scale=0.04]{orcidid.pdf}}}
\affiliation{Department of Physics and Materials Science, University of Luxembourg, L-1511 Luxembourg}
\author{Adolfo del Campo \href{https://orcid.org/0000-0003-2219-2851}{\includegraphics[scale=0.04]{orcidid.pdf}}}
\affiliation{Department of Physics and Materials Science, University of Luxembourg, L-1511 Luxembourg}
\affiliation{Donostia International Physics Center,  E-20018 San Sebasti\'an, Spain}

\author{Apollonas S. Matsoukas-Roubeas \href{https://orcid.org/0000-0001-5517-0224}{\includegraphics[scale=0.04]{orcidid.pdf}}}
\thanks{Corresponding authors:\\ pablo.martinez-azcona@tum.de, sam310@cam.ac.uk}
\affiliation{Department of Applied Mathematics and Theoretical Physics,
University of Cambridge, Cambridge CB3 0WA, United Kingdom}

\date{\today}

\begin{abstract}
Krylov subspace methods provide an efficient tool for modeling any quantum initial value problem as a one-dimensional chain. 
A time-ordered path formulation, in which the dynamical generator successively splits probability between paths, yields a speed limit within such dynamical subspaces.
For power-law growing Lanczos coefficients $b_n \propto  n^\delta$ we find super-ballistic lightcones, with an exponentially spreading lightcone for maximal growth $\delta = 1$. 
Causality in the Krylov chain constrains the Lanczos coefficients to grow at most linearly.
We illustrate the validity of the above in the $\mathsf{SL}(2, \mbb R)$ and Heisenberg-Weyl algebras, and in a many-body skew field Ising model. Moreover, we extend the above bound to the family of $\mc Q$-complexities.
Lastly, mapping to a spin chain gives a Lieb–Robinson bound analogue and opens the problem to standard condensed-matter methods.
\end{abstract}

\maketitle

The development of quantum mechanics in the last century \cite{Heisenberg1925, Schroedinger1926} has been intimately tied to diagonalization of the dynamical generator. It provides a powerful description of the time evolution of a quantum system, along with its possible measurement outcomes \cite{Born1925}. A complementary systematic representation of quantum dynamics, which considers the initial operator \footnote{In this letter, we focus primarily on the Heisenberg picture where the time evolution acts on the observable, though our results can be readily applied and are partially presented in the Schrödinger picture.}, is obtained by tridiagonalizing the dynamical generator \cite{Lanczos1950}.
Such a tridiagonalization is associated to a Krylov subspace constructed through repeated application of the dynamical generator  \cite{Krylov1931}. 
Orthogonalizing the resulting sequence via the Lanczos algorithm yields the Krylov basis  \cite{Viswanath1994}, on which the dynamical generator becomes a tridiagonal matrix given by the sequence of Lanczos coefficients $\{b_n\}$. 
In this representation, any quantum initial value problem is mapped onto a nearest-neighbor one-dimensional Krylov chain. 

Krylov methods have long been used in condensed matter physics and numerical analysis for their numerical efficiency \cite{Viswanath1994, liesenbook}. 
More recently, they have attracted renewed attention as a tool for studying the notion of complexity and characterizing properties of quantum dynamics \cite{parker_universal_2019, NANDY20251}.
In a many-body system, repeated application of the generator on a local initial operator yields increasingly non-local operators. Therefore, the first elements of the Krylov basis are simple, while later ones are more complex.  
This motivates the definition of the mean of the probability distribution in the Krylov chain, which was proposed as a measure of operator complexity, dubbed Krylov complexity. Such a measure has spurred extensive work in many-body and high-energy physics \cite{barbon_evolution_2019,avdoshkin_euclidean_2020,dymarsky_krylov_2021,rabinovici_krylov_2022,caputa_geometry_2022,balasubramanian_quantum_2022,hornedal_ultimate_2022,avdoshkin_krylov_2022,rabinovici_bulk_2023,Takahashi2024,Bhattacharjee2023,Grabarits2026KCQPT} and was summarized in several reviews \cite{NANDY20251,baiguera2025quantumcomplexitygravityquantum,rabinovici2025krylovcomplexity}. 
The universal operator growth hypothesis (UOGH) \cite{parker_universal_2019} states that the maximal growth of Lanczos coefficients is linear, $b_n = \alpha n$, which is saturated in certain chaotic models \cite{parker_universal_2019,NANDY20251} as well as systems exhibiting saddle-dominated scrambling \cite{Bhattacharjee_2022}. Here, we demonstrate that causality within the Krylov chain implies that the maximum growth of Lanczos coefficients must be linear.
 
Propagating deeper into the Krylov chain corresponds to longer times, when higher-order time derivatives become significant. The family of probability distributions over the Krylov chain at all times takes values over a plane $(n,t) \in \mathbb N \times \mathbb R_+$, where the information propagates in a way akin to the light cone of special relativity.
The Lieb-Robinson (LR) bound \cite{Lieb1972, Nachtergaele2006, Hastings2006} rigorously establishes that an effective light-cone emerges in non-relativistic local many-body quantum systems, in which information propagates at a finite speed \cite{nachtergaele_lieb-robinson_2010, kliesch_lieb-robinson_2014, anthony_chen_speed_2023}, and bounds Out-of-Time-Order-Correlators (OTOC's) \cite{roberts_lieb-robinson_2016, maldacena_bound_2016}. 
Other bounds such as quantum speed limits \cite{mandelstam_uncertainty_1991, margolus_maximum_1998, Gong2022, deffner_quantum_2017, delCampo_QSLopen_2013, shanahan_QSL_2018} constrain how fast a quantum system can transition between two states, and bound the growth of the Krylov complexity \cite{hornedal_ultimate_2022,Hornedal2023, Grabarits2026}.

\begin{figure}[t]
    \centering
    \vspace{8pt}
    \includegraphics[width=\linewidth]{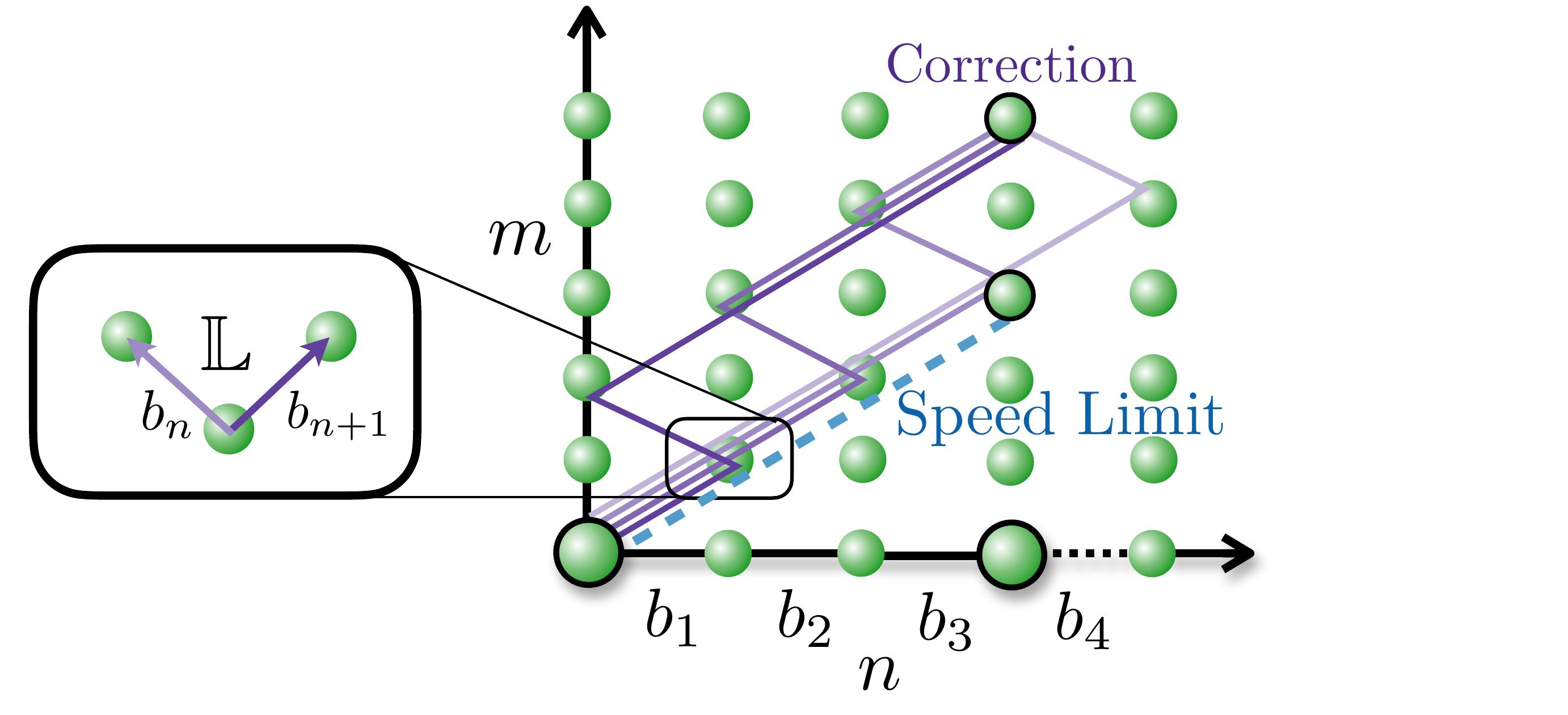}
    \caption{(a)\textbf{ Paths in the Krylov chain and bounds}. The paths are generated by the recurrence relation (inset). The two axes are $m$, corresponding to the power of the generator and time $\sum_m (i \mbb L t)^m$ and the position in the Krylov chain $n$. The speed limit (blue) is given by the direct path $\mathbb P_{0\to n}^n$ [cf. \eqref{eq:matrix_el_Krylov}]. The correction (violet) is given by the paths $\mathbb P_{0\to n}^{n+2}$.}
    \label{fig:illustrationKrylovLR}
\end{figure}

In this article, we formulate quantum dynamics using time-ordered paths in Krylov subspaces, where the generator divides probability across branching trajectories. This yields an explicit speed limit on information spread within these dynamical subspaces, governed by the shortest direct path from the origin to any site [cf. Fig. \ref{fig:illustrationKrylovLR}(a)].   
The sequence of Lanczos coefficients $\{b_n\}$ determines the shape of the light cone.
For the case of power-law growth $b_n \propto n^\delta$ we find a super-ballistic light cone, which converges to an exponentially spreading light cone in the $\delta \to 1$ limit.
We illustrate the regimes of validity of the bounds in the $\mathsf{SL}(2, \mbb R)$, Heisenberg-Weyl ($\mathsf{HW}$) complexity algebras \cite{caputa_geometry_2022}, and in the skew field Ising model.
Moreover, we show that our result bounds a family of $\mc Q$-complexities \cite{parker_universal_2019}.
Lastly, we provide a general mapping of the Krylov dynamics onto a many-body spin chain, through which we interpret our speed limit as a Lieb–Robinson bound.

The framework and bounds established here can be readily applied to state speed limits in the Schrödinger picture \cite{balasubramanian_quantum_2022} and connect naturally to other speed limits in time-dependent generalizations of Krylov subspace methods \cite{takahashi_krylov_2025}.

\textbf{\emph{Paths in the Krylov subspace}}---
Under the Heisenberg evolution generated by a Liouvillian $\mathcal{L}$, $\partial_t \hat{O}_t = i \mathcal{L}(\hat{O}_t)$, an initial operator $\hat{O}_0$ remains within the Krylov subspace $\mathcal{K}_{\mathcal{L}, \hat{O}_0} = \text{span}\{\mathcal{L}^n(\hat{O}_0)\}_{n=0}^{D_K-1}$, where $D_K$ is the dimension of the subspace.
It is standard to treat operators as vectors in a larger Hilbert space \cite{gyamfi2020}, $\hat{O}_0 \mapsto \vert{}O_0)$, such that the Liouvillian superoperator becomes a matrix $\mathcal{L}(\bullet) \mapsto \mathbb{L}$ \footnote{In this letter, we refer to the generator in the Heisenberg picture—technically the adjoint Liouvillian—simply as the Liouvillian.}.
In the orthonormal Krylov basis $\{|K_n)\}_{n=0}^{D_K -1}$ the Liouvillian becomes tridiagonal and reads
\begin{align}
\label{krylovchain}
      \mathbb L= \!\sum_{n=1}^{D_K-1}\!\!{b}_{n} \left(|K_{n-1})(K_n| +|K_{n})(K_{n-1}|\right)  ,
\end{align}
obeying the recurrence relation $\mbb L |K_n) = b_n |K_{n-1})+b_{n+1}|K_{n+1})$.
The dynamics is governed by the discrete Schr\"odinger equation
$\dot \varphi_n = b_n \varphi_{n-1} - b_{n+1}\varphi_{n+1}$, 
which dictates the evolution of the probability amplitudes $\varphi_n(t)$  \cite{parker_universal_2019}. The time-evolved operator in this basis is $|O_t) = \sum_{n=0}^{D_K-1} i^n \varphi_n(t) |K_n)$, 
subject to the initial condition $\varphi_n(0)=\delta_{n0}$, where $\delta_{ij}$ is the Kronecker delta.
The Krylov complexity is the average position on this chain $K(t) = \sum_{n=0}^{D_K-1} n |\varphi_n(t)|^2$.

The tridiagonal form of the dynamical generator allows the dynamics in the Krylov chain to be understood combinatorially in terms of paths \cite{avdoshkin_krylov_2022, Oste2015, Brak2009, NANDY20251, parker_universal_2019, balasubramanian_quantum_2022}. Let us consider the general case, starting from the site $j$, any application of the dynamical generator $\mbb L$ can introduce a jump to the previous site with weight $b_{j}$, 
or a jump to the next with weight $b_{j+1}$ [cf. Fig. \ref{fig:illustrationKrylovLR} (inset)].
Thus, any matrix element can be written as the sum over all paths connecting site $j$ to site $k$ in $m$ steps, denoted as the set $\mathbb P_{j\to k}^m\!=\!\{\bs p\! =\! (p_0, p_1, \cdots, p_m)| p_0\! = j,  p_m = k\}$,
\begin{equation}\label{eq:matrix_el_Krylov}
    (K_k|\mbb L^m|K_j) = \!\!\sum_{\bs p \in \mathbb P_{j\to k}^m} \! \! b_{p_1}  b_{p_2} \cdots  b_{p_m}.
\end{equation}
Quantum dynamics in this representation can be comprehended through a sum of possible paths, weighted by products of the Lanczos coefficients.

To connect the discrete path formulation to standard path approaches in quantum dynamics, one can construct an integral relation. In particular, removing the bond $\mathbb V = b_n \left( \vert{}K_n)(K_{n-1}\vert{} + \vert{}K_{n-1})(K_n\vert{} \right)$ decouples the chain between sites $n-1$ and $n$, defining a cut Liouvillian $\mathbb L_0 = \mathbb L - \mathbb V$ that splits the Krylov space into two non-interacting dynamical subspaces. By defining the Green's function $G^0_{nn}(t) \equiv (K_n\vert{} e^{-i \mathbb L_0 t} \vert{}K_n)$, one obtains the relation
$\varphi_n(t) = -i b_n \int_0^t G^0_{nn}(t-s) \varphi_{n-1}(s) ds$.
A derivation is provided in Appendix \ref{app:integral}.
One can then recursively substitute the previous amplitude into the integrand down to an arbitrary site $\ell$ (where $0 \le \ell < n$). 
By defining the time-ordering integration domain as the simplex $\Delta_{n-\ell}(t) = \{ (t_\ell, \dots, t_{n-1}) \mid 0 \le t_\ell \le \cdots \le t_{n-1} \le t \}$ with $t_n \equiv t$, the probability amplitude at site $n$ unfolds into a multiple integral
\begin{align} \label{eq:dyson}
\varphi_n(t) = (-i)^{n-\ell} \! \! \int_{\Delta_{n-\ell}(t)} \mspace{-25mu} \mr d \bs t  \; \;\prod_{j=\ell+1}^n \! \! \! \! b_j G^0_{jj}(\delta t_j)  \varphi_\ell(t_\ell) ,
\end{align}
As detailed below, this path-based approach is directly linked to causality arguments associated with the UOGH.

\textbf{\emph{A speed limit on the Krylov chain---}}
By leveraging the bounds $|\int f(x) \mr d x|\leq \int |f(x)|\mr d x$ and $|G^0_{jj}(t)|\leq 1$ and setting $\ell=0$, Eq.~\eqref{eq:dyson} reduces to a strict recursive inequality for the probability amplitudes 
\begin{equation}\label{eq:amplitude_bound}
    |\varphi_n(t)|\leq \frac{t^n}{n!} b_1 b_2 \cdots b_n \equiv \left(\frac{t}{\tau_n}\right)^n.
\end{equation}
We have introduced the characteristic time to reach the $n$-th site $\tau_n := \sqrt[n]{\frac{ n!}{b_1 b_2 \cdots b_n}}$.
See Appendix~\ref{app:bound} for a detailed derivation. Note that this bound sets an upper bound on the Krylov complexity $K(t) \leq \sum_n n (\frac{t}{\tau_n})^{2 n}$.
Setting a threshold $\epsilon$ on the $n$-th amplitude yields the speed limit on the time at which the amplitude equals that threshold 
\begin{equation}
    \label{eq:speedlim}
    t_\epsilon \geq \sqrt[n]{\epsilon} \tau_n.
\end{equation}
It characterizes the minimum propagation time of information in the Krylov chain.

In the case that the Lanczos coefficients follow a power law  $b_n = \alpha n^\delta$, we find the time-scale to be $
    \alpha \tau_n =  n!^{\frac{1-\delta}{n}}$, which saturates to unity for linearly growing Lanczos coefficients $\delta = 1$.
 For this case, the bound can be written in a way akin to the lightcone of the Lieb-Robinson bound. In order to do this, we use the bounds $\log(x)\leq -(1-x)$ and $(\frac{n}{e})^n\leq n!$ and find $|\varphi_n(t)|\leq (n!)^\delta \frac{( \alpha t)^n}{n!} \leq e^{ - n \left( 1 - \frac{e^{1-\delta} \alpha t}{n^{1-\delta}}\right)}$. 
From the upper bound we can find a non-linear light-cone (LC) $v_\mathsf{a} \tau^\mr{LC}_n = n^{1-\delta} $ where $v_\mathsf{a} = \frac{e^{1-\delta}\alpha}{1 + \mathsf{a} \frac{\log \epsilon}{n}}$. The choice of the parameter $\mathsf a = 0$ corresponds to choosing the light-cone at the point that the exponent vanishes, thus having no notion of $\epsilon$, while $\mathsf{a}=1$ corresponds to setting the amplitude equal to $\epsilon$. Note that for fixed $\epsilon$ and large $n$ the latter converges asymptotically to the former $v_1 \to v_0$. In the asymptotic limits of large $n$ and $\delta \to 1$ it is possible to justify that the power $n \sim t^{1/(1-\delta)} \to e^{t}$ converges to an exponentially spreading light-cone.

The physics of the speed limit only involves the direct path [cf. Fig. \ref{fig:illustrationKrylovLR}~(a) blue], but one can naturally compute the next order of the expansion $\varphi_n(t) = t^n \frac{(K_n|\mbb L^n|K_0)}{n!} + t^n \frac{(K_n|\mbb L^{n+2}|K_0)}{(n+2)!} + \mc O(t^{n+4})$ from the sum of all paths $\mbb P_{0\to n}^{n+2}$ connecting $0 \to n$ in $n+2$ steps. 
Each path involves the product of the Lanczos $b_1 \cdots b_n$ with an additional $b_k^2$ coming from adding one ``fold'' to the path [cf. Fig. \ref{fig:illustrationKrylovLR} (a) (violet lines)]. Solving for the time at which the next order expansion equals the threshold gives [cf. App. \ref{sec:longDyckpath}]
\begin{equation}
\label{eq:corr_speed_limit}
    \tau^\mr{cSL}_n \approx \sqrt[n]{\epsilon}\tau_n\left(1 + \frac{ (\sqrt[n]{\epsilon} \tau_n)^2}{n (n+1) (n+2)} \sum_{k=1}^{n+1} b_k^2 \right).
\end{equation}
The corrected speed limit (cSL) is larger than the direct-path speed limit (SL), $\tau^\mr{cSL}_n\geq \tau^\mr{SL}_n$ and, as such, does not necessarily set a lower bound to $ t_\epsilon$.
Rather, we expect $\tau^\mr{cSL}_n$ to provide a tighter estimate of the actual numerical value of $ t_\epsilon$ [cf. App.~\ref{sec:sm_num_efficiency}].

\emph{\textbf{Causality bounds the Lanczos coefficients}}--- 
The time-scale of propagation to a given $n$-th site $\tau_n$ allows for a natural definition of causality in Krylov space. For propagation to be causal, the time-scale $\tau_n$ should grow monotonically with $n$, i.e., $\tau_{n-1} \leq \tau_n$. This implies that information, initially localized in $|K_0)$, reaches lower sites in the Krylov chain first. Equivalently, the site $|K_{n+1})$ cannot be reached before  $|K_n)$. 
In the context of the UOGH \cite{parker_universal_2019} causality implies that the information propagates from simpler operators at lower $n$, towards more complex operators at larger $n$.
The condition $\tau_{n-1}\leq \tau_{n}$ can be written as $ ( \tau_{n-1})^{n}\leq \frac{n}{b_{n}} (\tau_{n-1})^{n-1}$ which gives a bound on the Lanczos coefficient $b_n$ that depends on the time-scale of reaching the previous site $\tau_{n-1}$
\begin{equation}\label{eq:causality_bound}
    b_n \leq \frac{n}{ \tau_{n-1}} = n \sqrt[n-1]{\frac{b_1 \dots b_{n-1}}{(n-1)!}}.
\end{equation}
When $\tau_{n-1}$ is an increasing function of $n$ the causality bound imposes a tighter constraint on $b_n$, which is now forced to be sublinear in $n$. By causality, we also have a whole family of looser inequalities $b_n\leq \frac{n}{\tau_j}$ with the time to reach any of the previous sites $j< n-1$.
These observations connect the Krylov light cone to the UOGH \cite{parker_universal_2019}. Simply by taking the case $j=1$ in the previous inequality we find
\begin{equation}\label{eq:ogh_bound}
    b_n \leq b_1\, n,
\end{equation}
so that in a causal Krylov chain the Lanczos coefficients grow at most linearly, with a slope upper bounded by the first Lanczos coefficient, which also sets the bound for the growth of Krylov complexity \cite{hornedal_ultimate_2022}. This implies that $\alpha \leq b_1$ and thus it also imposes a bound on the Lyapunov exponent $\lambda \leq 2 \alpha \leq 2 b_1$ \cite{parker_universal_2019}.
The maximal linear growth was previously derived from the high-frequency decay of the spectral function \cite{parker_universal_2019}.
Here, it follows from the direct path alone.
We refer the reader to App. \ref{sec:sm_causality} for a detailed discussion on causality in the Krylov chain.

 \begin{figure}[t]
    \centering
    \includegraphics[width=\linewidth]{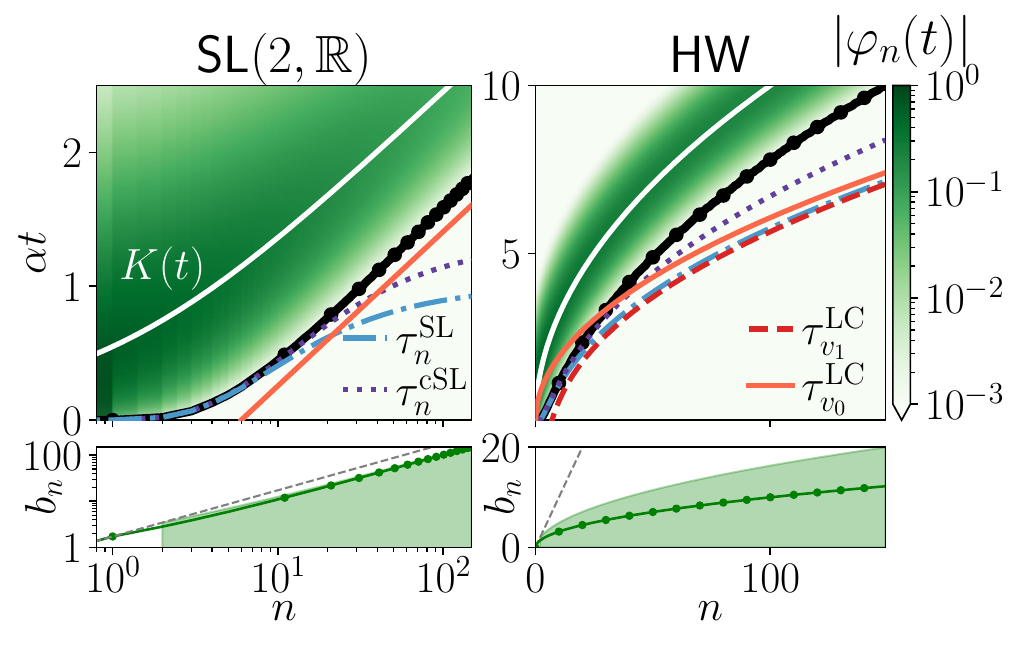}
    \caption{\textbf{Dynamics in the Krylov chain and bounds for the complexity algebras}: (left) $\mathsf{SL}(2, \mathbb R)$ with $\eta = 3$ and (right) Heisenberg-Weyl. The plots show the threshold $ t_\epsilon$ (black solid line) where we choose $\epsilon = 10^{-3}$ in Eq.~\eqref{eq:speedlim}, the SL (dash-dotted blue), the corrected SL (dotted purple) and the lightcone (red) with $\mathsf a=0$ (thin solid line) and with $\mathsf a =1$ (thick dashed line). We also show the Krylov complexity $K(t)$ (white). (down) Lanczos coefficients (green line dotted every 10 points) and causality bound on $b_n$ \eqref{eq:causality_bound} (shaded green area) along with the UOGH bound \eqref{eq:ogh_bound} (dashed gray line). $\mathsf{SL}(2, \mbb R)$ has a logarithmic scale on $n$ to highlight the exponentially spreading light cone.}
    \label{fig:dynamics_comp-algs}
\end{figure}
\textbf{\emph{Complexity algebras}}---
We now consider the closed complexity algebras \cite{caputa_geometry_2022} where $b_n$ and $\varphi_n(t)$ are known analytically.
Here we will consider $\mathsf{SL}(2, \mathbb R)$
with asymptotically linear Lanczos coefficients $b_n = \alpha \sqrt{n(n-1+\eta)}$, closely related to the Sachdev-Ye-Kitaev model \cite{parker_universal_2019, NANDY20251}, and the Heisenberg-Weyl algebra ($\mathsf{HW}$), which has sublinear Lanczos coefficients $b_n = \alpha \sqrt{n}$. [cf. \cite{caputa_geometry_2022, NANDY20251} for the analytical expressions of $\varphi_n(t)$]
Figure \ref{fig:dynamics_comp-algs} shows the evolution of the probability amplitudes $|\varphi_n(t)|$ (color scale), from which we compute the time $ t_\epsilon$ numerically (solid black lines). We show the SL (blue dash dotted), its correction cSL (purple dotted), and the asymptotic light cone (red) with the two choices for the speed $v_0$ (solid thin line), and $v_1$ (dashed thick line). The former has no notion of $\epsilon$ and cannot be directly compared with the rest of the bounds. 

Figure \ref{fig:dynamics_comp-algs} (left) shows that the SL (blue dash-dotted) is tight at short times and saturate to a constant at large $n$.
The correction $\tau^{\rm cSL}_n$  (purple dotted) is tighter than the SL, although it still saturates to a constant at long times.
In contrast, the $\delta \to 1$ limit of the light cone gives the right behavior at long times $n \sim e^{2 \tau^\ts{lc}}$ with the exponentially spreading light-cone.
The plot of the Lanczos coefficients (lower) shows that in the case of maximal linear growth, we are close to saturating the bound imposed by causality \eqref{eq:causality_bound}, and in addition this bound is close to the universal linear bound \eqref{eq:ogh_bound} (dashed gray line). 
Fig.~\ref{fig:dynamics_comp-algs}~(right) shows that for $\mathsf{HW}$ the SL is larger due to the slower growth of the Lanczos. The SL now does not saturate to a constant at long times, but keeps growing with $n$. The correction, is tighter than the SL, although it still has a notable difference to the numerical time. We find that both versions of the light-cone (with and without knowledge of $\epsilon$, $\mathsf a = 0, 1$, red lines) are comparably tight to the SL.

\begin{figure}[t]
    \centering
    \includegraphics[width=\linewidth]{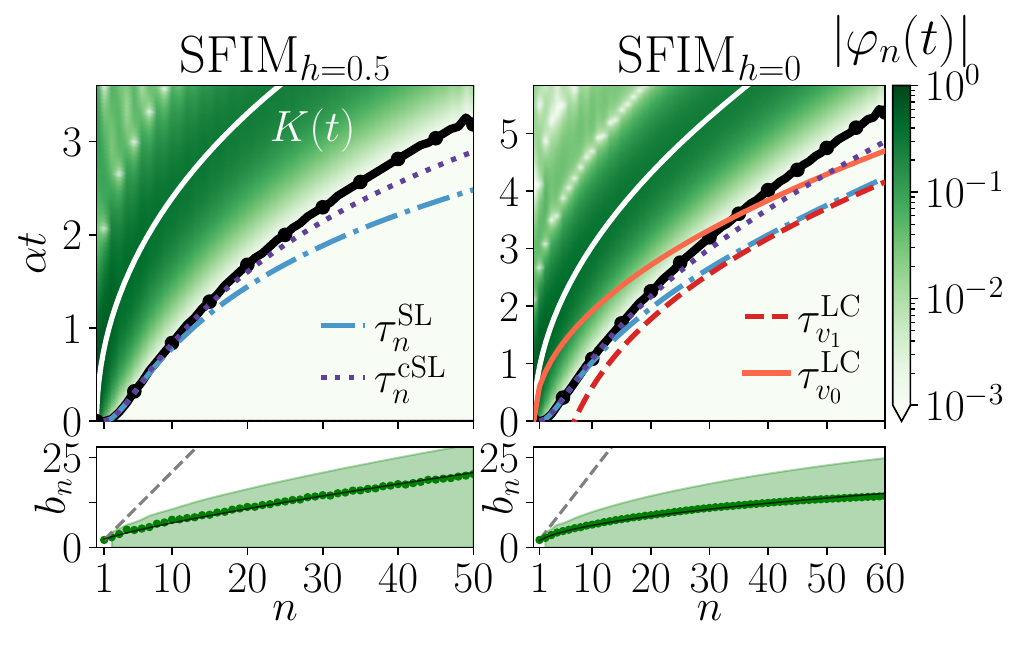}
    \caption{\textbf{Dynamics in the Krylov chain and bounds for Skew Field Ising model} with $L=50$ spins in the chaotic $h=0.5$ (left) and integrable $h=0$ (right) phases. (lower insets) Lanczos coefficients (green dots), with their asymptotic $b_n =\alpha n/\log n$ (left) and $b_n = \alpha \sqrt{n}$ (right) and the causality bound (green shaded area) along with the UOGH bound \eqref{eq:ogh_bound} (dashed gray line). The bounds depicted are the same as in Fig. \ref{fig:dynamics_comp-algs}: Light cones with $v_0, v_1$ (right), speed limit (blue dash dotted) and correction to the speed limit (purple dotted). We also show the Krylov complexity $K(t)$ (white)}
    \label{fig:SFIM_long}
\end{figure}

\textbf{\emph{Skew-Field Ising model}}---
We now consider a paradigmatic many-body quantum system, the skew-field Ising model (SFIM) given by the Hamiltonian 
\begin{equation}
    \label{eq:SFIM}
    \hat H = \sum_{j=1}^L (\hat \sigma^x_j \hat \sigma^x_{j+1} + h \hat \sigma_j^x +h_z \hat \sigma^z_j).
\end{equation}
We compute the first $\mu$ Lanczos coefficients $\{b_n\}_{n=1}^{\mu}$ using the $\texttt{PauliStrings.jl}$ package \cite{Loizeau2025, Loizeau2025_code} for SFIM with $L=50$ and fixed $h_z=- 1.05$, and an initial operator $\ha O_0 = \ha \sigma^x_L$, both in the chaotic $h=0.5$ and integrable $h=0$ phases, with  $\mu=50$ and $\mu = 60$ sites respectively. For an analysis of the full Krylov chain dynamics in a short SFIM, see Appendix [cf. App. \ref{sec:dynamics_shortSFIM}].
Figure \ref{fig:SFIM_long} shows the dynamics in the Krylov chain of these two models, with Lanczos $b_n = \frac{\alpha n}{\log(n)}$ and $b_n = \alpha \sqrt n$. As expected, SL and cSL are tighter at short times while the light cone is tighter asymptotically.
The plot of the Lanczos coefficients (lower) shows that these models are not close to saturating the causality bound, with the chaotic phase being closer to the upper bound. 

\textbf{\emph{Bound on the $\mc Q$-complexities}}---
Krylov complexity bounds a whole family of $\mc Q$-complexities \cite{parker_universal_2019}.
$\mc Q=\sum_a q_a |q_a)(q_a|$ is any positive semidefinite superoperator for which there exists a constant $M>0$ such that $(q_a|\mc L|q_b)=0, \, (q_a|O_0)=0$ for $|q_a-q_b|>M$ and $|q_a|>M$, respectively. These conditions imply that the $\mc Q$-eigenbasis has a bounded number of nonzero components on the Krylov basis, i.e., $(q_a|K_n)=0$ for $q_a > 2 M n$. We now find a refined bound on the projection on each $\mc Q$-eigenstate based on the Krylov LR bound [cf. App. \ref{sm:qcomp}]
\begin{equation}
    |(q|O_t)|^2  \lesssim \left(\frac{t}{\tau_{\lceil \frac{q}{2M}\rceil} }\right)^{2 \lceil \frac{q}{2M}\rceil}. 
\end{equation}
Therefore, we see that the LR bound derived here applies to quantities beyond the Krylov chain and beyond the average. 
Two notable members of this family are \cite{parker_universal_2019}: the operator-size superoperator, $\mc Q |X^{\bs{\nu}}) = \mr{size}(\bs \nu)|X^{\bs{\nu}})$, where $X^{\bs{\nu}}$ is a Pauli string and $\mr{size}(\bs \nu)$ counts its non-zero components; and OTOCs, for which $(A_t|\mc Q|B_t) = \sum_i([V_i, A_t]|[V_i, B_t])$.

\textbf{\emph{Spin chain representation of Krylov evolution---}} 
In order to formalize the previous SL through a Lieb-Robinson bound, we now construct a spin chain that mimics the Krylov chain. We set the length equal to the Krylov dimension $L = D_K$, 
and the nearest-neighbor coupling to be the Lanczos coefficients $b_n$ as
\begin{align}
\label{spinchain}
     \ha H\ep{\mr{sc}} \! = \! \frac{1}{2}\sum_{n=1}^{D_K-1}\! 
     b_n (\ha \sigma^x_{n-1}\ha \sigma^x_n + \ha \sigma^y_{n-1} \ha \sigma^y_n) .
\end{align}
The single excitation sector of this Hamiltonian gives the Krylov chain when we consider the initial state in the first site $\ket{\uparrow_0}=\ket{\uparrow \downarrow \downarrow \cdots}$, Note that $[\ha H\ep{\mr{sc}}, \sum_n \ha \sigma^z_n]=0$ and as such the total number of excitations is conserved. 
The amplitudes at site $n$ in this spin chain evolve in the same way as the probability amplitudes [cf. App. \ref{sec:sm_spinchain}] through the discrete Schrödinger equation and, importantly, do not depend explicitly on the Krylov basis vectors, which are only required to map the chain dynamics to the original Hilbert space.
The mapping endows the Krylov space with a tensor product structure promoting $\msc K \cong \mbb C^{D_K}$ to $\msc K^\mr{sc} \cong (\mbb C^2)^{\otimes D_K} $, which has a spin $\frac{1}{2}$ degree of freedom at each site of the chain, while the single-excitation manifold has dimension $D_K$ by construction.

The above allows one to apply tools from quantum many-body physics directly to the introduced spin chain, and directly justifies why a Lieb-Robinson type bound applies to the Krylov chain, albeit different to the ones derived in \cite{matsuura2026krylovbreaktimesinhomogeneous, alishahiha2026}. However, the bound \eqref{eq:amplitude_bound} is tighter than the one obtained in this approach.
We highlight that any quantum initial value problem can be simulated on a spin-$\frac{1}{2}$ nearest-neighbor hopping chain, provided the hoppings can be tuned to match the desired Lanczos coefficients, which can be engineered in many current quantum platforms. This mapping allows to simulate Krylov complexity, and related quantities, directly in current quantum experiments.

\textbf{\emph{Conclusion}}---
By employing a path-based approach to quantum dynamics in Krylov space, we have shown that probability amplitudes satisfy strict recursive inequalities. This establishes a fundamental speed limit and light cone for information propagation within the dynamical subspaces of arbitrary isolated many-body quantum systems. Crucially, imposing an inherent notion of causality through time-ordering mandates that Lanczos coefficients grow at most linearly ($b_n \le b_1 n$), providing a mathematically simple origin for the UOGH. We verified these analytical limits in the $\mathsf{SL}(2,\mathbb{R})$ and Heisenberg-Weyl complexity algebras, along with the many-body skew-field Ising model. Beyond tracking mean operator complexity, the derived constraints also apply to the broader family of $Q$-complexities. Finally, by translating abstract Krylov evolution into a physical spin-1/2 hopping chain, we established an analogue to the Lieb-Robinson bound. Directly linking to condensed-matter methods, this correspondence provides a concrete scheme for simulating Krylov subspace dynamics on existing platforms.


\emph{Acknowledgements}---
We thank Budhaditya Bhattacharjee, Aurélia Chenu, Niklas Hörnedal, Neill Lambert,  Peter Rabl, Ulrich Schneider, and Ruth Shir for insightful discussions. 
P.M.A. acknowledges
funding from the Swiss National Science Foundation (Project No. CRSII 222812/1). This research
is part of the Munich Quantum Valley initiative,
which is supported by the Bavarian state government with funds from the Hightech Agenda Bayern Plus. This work was partially funded by the Luxembourg National Research Fund (FNR, Attract grant 15382998).
A.S.M.-R. was supported by the EPSRC UK Multidisciplinary Centre for Neuromorphic Computing (grant UKRI982).


%

\pagebreak 
\clearpage
\newpage

\onecolumngrid
\appendix

\section{The integral equation of amplitudes} \label{app:integral}
Here, we derive an integral equation that serves as an analogue to the Dyson equation on the Krylov chain—or, more generally, a Volterra convolution equation. 
Consider the Krylov basis states $\{\vert{}K_0), \vert{}K_1), \dots\}$ and the Liouvillian \eqref{krylovchain}
\begin{align}
\mathbb L= \sum_{n=1}^{D_K-1} b_{n} \left( |K_{n-1})(K_n| +|K_{n})(K_{n-1}| \right)  .
\end{align}
Let us then define a cut Liouvillian without the bond connecting site $n-1$ and site $n$
\begin{equation}
   \mathbb L_0 = \mathbb L - \mathbb V \;, \quad \text{where} \quad \mathbb V = b_n \Big( \vert{}K_n)(K_{n-1}\vert{} + \vert{}K_{n-1})(K_n\vert{} \Big) ,
\end{equation}
which decouples the chain into two dynamical subspaces
$\msc K_{\mc L, \ha O_0}^{(A)} = \text{span} \{\vert{}K_0), \dots, \vert{}K_{n-1}) \}$ 
and
$\msc K_{\mc L, \ha O_0}^{(B)} = \text{span} \{\vert{}K_n), \vert{}K_{n+1}), \dots \}$.
Let us also consider the function $F(s) = e^{i \mathbb L_0 s} e^{-i \mathbb L s}$. 
Differentiating with respect to $s$
\begin{equation*}
\frac{d}{ds} F(s) = i \mathbb L_0 e^{i \mathbb L_0 s} e^{-i \mathbb L s} + e^{i \mathbb L_0 s} (-i \mathbb L) e^{-i \mathbb L s} = -i e^{i \mathbb L_0 s} (\mathbb L - \mathbb L_0) e^{-i \mathbb L s} = -i e^{i \mathbb L_0 s} \mathbb V e^{-i \mathbb L s} .
\end{equation*}
Integrating both sides from $s = 0$ to $s = t$ gives
\begin{equation*}
F(t) - F(0) = \int_0^t \frac{d}{ds} F(s) \, ds \implies e^{i \mathbb L_0 t} e^{-i \mathbb L t} - I = -i \int_0^t e^{i \mathbb L_0 s} \mathbb V e^{-i \mathbb L s} \, ds .
\end{equation*}
Multiplying from the left by $e^{-i \mathbb L_0 t}$ yields Duhamel's exact integral equation
\begin{equation}
\label{duhamel}
e^{-i \mathbb L t} = e^{-i \mathbb L_0 t} - i \int_0^t e^{-i \mathbb L_0 (t-s)} \mathbb V e^{-i \mathbb L s} \, ds .
\end{equation}

Applying the initial operator, $\vert{}K_0)$ on the right and project onto $(K_n\vert{}$ on the left of eq.~\eqref{duhamel} gives
\begin{equation*}
(K_n\vert{} e^{-i \mathbb L t} \vert{}K_0) = (K_n\vert{} e^{-i \mathbb L_0 t} \vert{}K_0) - i \int_0^t (K_n\vert{} e^{-i \mathbb L_0 (t-s)} \mathbb V e^{-i \mathbb L s} \vert{}K_0) \, ds .
\end{equation*}
Since $\vert{}K_0) \in \msc K_{\mc L, \ha O_0}^{(A)}$ and $\mathbb L_0$ does not couple $\msc K_{\mc L, \ha O_0}^{(A)}$ to $\msc K_{\mc L, \ha O_0}^{(B)}$ dynamically, the evolved operator $e^{-i \mathbb L_0 t} \vert{}K_0)$ remains strictly in $\msc K_{\mc L, \ha O_0}^{(A)}$. 
Because $\vert{}K_n) \in \msc K_{\mc L, \ha O_0}^{(B)}$, the operators are orthogonal, $(K_n\vert{} e^{-i \mathbb L_0 t} \vert{}K_0) = 0$.

We now substitute $\mathbb V$ into the integrand
\begin{equation*}
(K_n\vert{} e^{-i \mathbb L_0 (t-s)} \mathbb V e^{-i \mathbb L s} \vert{}K_0) = b_n (K_n\vert{} e^{-i \mathbb L_0 (t-s)} \vert{}K_n) (K_{n-1}\vert{} e^{-i \mathbb L s} \vert{}K_0) + b_n (K_n\vert{} e^{-i \mathbb L_0 (t-s)} \vert{}K_{n-1}) (K_n\vert{} e^{-i \mathbb L s} \vert{}K_0) 
\end{equation*}
Again, since $\vert{}K_{n-1}) \in \msc K_{\mc L, \ha O_0}^{(A)}$, the time evolved $e^{-i \mathbb L_0 (t-s)} \vert{}K_{n-1})$ remains in $\msc K_{\mc L, \ha O_0}^{(A)}$, making $(K_n\vert{} e^{-i \mathbb L_0 (t-s)} \vert{}K_{n-1}) = 0$. 
Finally, we define the diagonal amplitude on site $n$ under the cut Liouvillian, which becomes our Green function, $G^0_{nn}(t) \equiv (K_n\vert{} e^{-i \mathbb L_0 t} \vert{}K_n)$, and recall that by definition the probability amplitude to find the state at site $n$ at time $t$ under full evolution starting from $\vert{}K_0)$ is $\varphi_n(t) \equiv (K_n\vert{} e^{-i \mathbb L t} \vert{}K_0)$.
Using the definitions for the remaining non-zero terms and substituting into the projected integral identity above
\begin{equation}\label{eq:int_equation_phin}
\varphi_n(t) = -i b_n \int_0^t G^0_{nn}(t-s) \varphi_{n-1}(s) \, ds .
\end{equation}

\section{Bounding the probability amplitudes} \label{app:bound}
In this appendix, we derive the bound of the probability amplitudes given in the main text.
We begin with the exact integral equation~\eqref{eq:int_equation_phin} for the amplitude $\varphi_n(t)$ derived above. Taking the absolute value of both sides, and noting that the absolute value of an integral is strictly less than or equal to the integral of the absolute values, we obtain
\begin{equation*}
\vert{}\varphi_n(t)\vert{} \le b_n \int_0^t \vert{}G^0_{nn}(t-s)\vert{} \vert{}\varphi_{n-1}(s)\vert{} , ds .
\end{equation*}
Note that the Lanczos coefficient $b_n$ remains outside the absolute value, as it is strictly positive by definition.
Also, since $G^0_{nn}(t) \equiv (K_n\vert{} e^{-i \mathbb L_0 t} \vert{}K_n)$ represents a transition amplitude under the Hermitian evolution governed by the cut Liouvillian $\mathbb L_0$, it is bounded by unity, $\vert{}G^0_{nn}(t-s)\vert{} \le 1$. Substituting this rule allows us to drop the Green function term, yielding the strict upper bound
\begin{equation}\label{eq:recursive_bound}\vert{}\varphi_n(t)\vert{} \le b_n \int_0^t \vert{}\varphi_{n-1}(s)\vert{} , ds .\end{equation}

Let us evaluate this bound iteratively, starting from the beginning of the Krylov chain. By definition, the system is initialized at site $0$, meaning that the maximum possible probability amplitude at site $0$ at any given time cannot exceed $1$. Therefore, we have $\vert{}\varphi_0(t)\vert{} \le 1$.

For $n=1$, applying Eq.~\eqref{eq:recursive_bound} gives\begin{equation*}\vert{}\varphi_1(t)\vert{} \le b_1 \int_0^t \vert{}\varphi_0(s_1)\vert{} , ds_1 \le b_1 \int_0^t 1 , ds_1 = b_1 t .\end{equation*}
For $n=2$, we substitute the bound we just found for $\vert{}\varphi_1(s_2)\vert{}$\begin{equation*}\vert{}\varphi_2(t)\vert{} \le b_2 \int_0^t \vert{}\varphi_1(s_2)\vert{} , ds_2 \le b_2 \int_0^t (b_1 s_2) , ds_2 = b_1 b_2 \frac{t^2}{2} .\end{equation*}
Proceeding to $n=3$, we substitute the bound for $\vert{}\varphi_2(s_3)\vert{}$\begin{equation*}\vert{}\varphi_3(t)\vert{} \le b_3 \int_0^t \vert{}\varphi_2(s_3)\vert{} , ds_3 \le b_3 \int_0^t \left(b_1 b_2 \frac{s_3^2}{2}\right) , ds_3 = b_1 b_2 b_3 \frac{t^3}{3!} .\end{equation*}
Generalizing to arbitrary $n$, we observe a clear recursive pattern. Each time we integrate a term of the form $s^k / k!$, we obtain $s^{k+1} / (k+1)!$. After $n$ nested integrations down the chain, we arrive at the final analytical bound for the probability amplitudes
\begin{equation}\vert{}\varphi_n(t)\vert{} \le b_1 b_2 \cdots b_n \frac{t^n}{n!} .
\end{equation}

\section{Contributions from longer Dyck paths}
\label{sec:longDyckpath}
\subsection{Vanishing diagonal terms}
When the diagonal Lanczos coefficients vanish, $a_k=0$, the first correction to the overlap appears two orders higher than the leading term,
\begin{equation}
    (K_n|{e^{-i \ha H t}}|O_0) = \frac{(-i t)^n}{n!}\, b_1 b_2 \cdots b_n + \frac{(-i t)^{n+2}}{(n+2)!} (K_n|{\mc L^{n+2}}|O_0) + \mc O(t^{n+4}).
\end{equation}
In path language, $(K_n|{\mc L^{n+2}}|O_0)$ is the weighted sum over all paths on the Krylov chain that connect site $0$ to site $n$ in $n+2$ steps. Such paths contain exactly one backward step; if this step retraces the bond between sites $k-1$ and $k$, the path acquires an additional weight $b_k^2$. Hence,
\begin{equation}
    (K_n|{\mc L^{n+2}}|O_0) = b_1 b_2 \cdots b_n \sum_{k=1}^{n+1} b_k^2,
\end{equation}
as illustrated below for $n=4$.

\begin{center}
\begin{tikzpicture}[scale=0.4,>=stealth]
  \draw[->] (-0.5,0) -- (5.5,0) node[below]{$n$};
  \draw[->] (0,-0.5) -- (0,6.5) node[left]{$t$};

  \draw[step=1cm,gray!20,very thin] (0,0) grid (5,6);

  \draw[thick,blue] 
    (0,0) -- (1,1) -- (2,2) -- (3,3) -- (4,4) -- (5,5) -- (4, 6);

  \foreach \x/\y in {0/0,1/1,2/2,3/3,4/4,5/5,4/6}{
    \filldraw[blue] (\x,\y) circle (2pt);
  }

  \node at (0.1,0.7) {$b_1$};
  \node at (1.1,1.7) {$b_2$};
  \node at (2.1,2.7) {$b_3$};
  \node at (3.1,3.7) {$b_4$};
  \node at (4.1,4.7) {$b_5$};
  \node at (4.1,5.7) {$b_5$};

\end{tikzpicture}
\begin{tikzpicture}[scale=0.4,>=stealth]
  \draw[->] (-0.5,0) -- (5.5,0) node[below]{$n$};
  \draw[->] (0,-0.5) -- (0,6.5) node[left]{$t$};

  \draw[step=1cm,gray!20,very thin] (0,0) grid (5,6);

  \draw[thick,blue] 
    (0,0) -- (1,1) -- (2,2) -- (3,3) -- (4,4) -- (3,5) -- (4, 6);

  \foreach \x/\y in {0/0,1/1,2/2,3/3,4/4,3/5,4/6}{
    \filldraw[blue] (\x,\y) circle (2pt);
  }

  \node at (0.1,0.7) {$b_1$};
  \node at (1.1,1.7) {$b_2$};
  \node at (2.1,2.7) {$b_3$};
  \node at (3.1,3.7) {$b_4$};
  \node at (4.1,4.7) {$b_4$};
  \node at (4.1,5.7) {$b_4$};

\end{tikzpicture}
\begin{tikzpicture}[scale=0.4,>=stealth]
  \draw[->] (-0.5,0) -- (5.5,0) node[below]{$n$};
  \draw[->] (0,-0.5) -- (0,6.5) node[left]{$t$};

  \draw[step=1cm,gray!20,very thin] (0,0) grid (5,6);

  \draw[thick,blue] 
    (0,0) -- (1,1) -- (2,2) -- (3,3) -- (2,4) -- (3,5) -- (4, 6);

  \foreach \x/\y in {0/0,1/1,2/2,3/3,2/4,3/5,4/6}{
    \filldraw[blue] (\x,\y) circle (2pt);
  }

  \node at (0.1,0.7) {$b_1$};
  \node at (1.1,1.7) {$b_2$};
  \node at (2.1,2.7) {$b_3$};
  \node at (3.1,3.7) {$b_3$};
  \node at (3.1,4.7) {$b_3$};
  \node at (4.1,5.7) {$b_4$};

\end{tikzpicture}
\begin{tikzpicture}[scale=0.4,>=stealth]
  \draw[->] (-0.5,0) -- (5.5,0) node[below]{$n$};
  \draw[->] (0,-0.5) -- (0,6.5) node[left]{$t$};

  \draw[step=1cm,gray!20,very thin] (0,0) grid (5,6);

  \draw[thick,blue] 
    (0,0) -- (1,1) -- (2,2) -- (1,3) -- (2,4) -- (3,5) -- (4, 6);

  \foreach \x/\y in {0/0,1/1,2/2,1/3,2/4,3/5,4/6}{
    \filldraw[blue] (\x,\y) circle (2pt);
  }

  \node at (0.1,0.7) {$b_1$};
  \node at (1.1,1.7) {$b_2$};
  \node at (2.1,2.7) {$b_2$};
  \node at (2.1,3.7) {$b_2$};
  \node at (3.1,4.7) {$b_3$};
  \node at (4.1,5.7) {$b_4$};

\end{tikzpicture}
\begin{tikzpicture}[scale=0.4,>=stealth]
  \draw[->] (-0.5,0) -- (5.5,0) node[below]{$n$};
  \draw[->] (0,-0.5) -- (0,6.5) node[left]{$t$};

  \draw[step=1cm,gray!20,very thin] (0,0) grid (5,6);

  \draw[thick,blue] 
    (0,0) -- (1,1) -- (0,2) -- (1,3) -- (2,4) -- (3,5) -- (4, 6);

  \foreach \x/\y in {0/0,1/1,0/2,1/3,2/4,3/5,4/6}{
    \filldraw[blue] (\x,\y) circle (2pt);
  }

  \node at (0.1,0.7) {$b_1$};
  \node at (1.1,1.7) {$b_1$};
  \node at (1.1,2.7) {$b_1$};
  \node at (2.1,3.7) {$b_2$};
  \node at (3.1,4.7) {$b_3$};
  \node at (4.1,5.7) {$b_4$};
\end{tikzpicture}
\end{center}

The arrival time at threshold $\epsilon$ is therefore determined by
\begin{equation}
    \epsilon = \frac{t^n}{n!}\, b_1 \cdots b_n \left( 1 - \frac{t^2}{(n+1)(n+2)} \sum_{k=1}^{n+1} b_k^2\right).
\end{equation}
Defining
\begin{equation}
    \delta \equiv \frac{\epsilon\, n!}{b_1 \cdots b_n}, \qquad Q \equiv \frac{1}{(n+1)(n+2)}\sum_{k=1}^{n+1} b_k^2,
\end{equation}
this becomes
\begin{equation}
    \delta = t^n \left( 1- Q t^2\right).
\end{equation}
To leading order the solution is $\tau = \delta^{1/n}$. To obtain the correction to the smallest positive root, we introduce $y = t/\tau$ and $\eta \equiv \tau^2$, so that
\begin{equation}
    1 = y^n \left(1 - Q \eta\, y^2\right).
\end{equation}
For $Q\eta \ll 1$, we expand $y = 1 + c\, \eta + \mc O(\eta^2)$ and find
\begin{equation}
    1 = \left(1 + c n \eta\right)\left(1 - Q \eta \right) + \mc O(\eta^2) \quad\Rightarrow\quad c = \frac{Q}{n}.
\end{equation}
The corrected arrival time thus reads
\begin{equation}
    \tau_{\mr{corr}} \approx \tau \left(1 + \frac{Q}{n}\,\tau^2\right), \qquad \tau = \delta^{1/n},
\end{equation}
corresponding to Eq.~\eqref{eq:corr_speed_limit}.

\subsection{Non-zero diagonal terms}
In the cases where the initial operator is non-Hermitian, or the Krylov subspace is constructed for quantum states in Schrödinger picture, there are non-vanishing diagonal elements $a_k$, which contribute already at order $t^{n+1}$.
For this reason, in this section, we write the initial pure state $\ket{\psi_0}$ and Krylov space basis vectors $\ket{K_n}$ using the standard Braket notation.
The paths of $n+1$ steps contain a single level step at site $k$ with weight $a_k$, while the paths of $n+2$ steps contain a backward step or two level steps at sites $j \leq k$. Collecting these contributions,
\begin{equation}
    \mel{K_n}{e^{-i \ha H t}}{\psi_0} = \frac{(-i t)^n}{n!}\, b_1 b_2 \cdots b_n \left( 1 - i q t - p t^2 + \mc O(t^3) \right),
\end{equation}
with
\begin{equation}
    q = \frac{1}{n+1}\sum_{k=0}^n a_k, \qquad p = \frac{1}{(n+1)(n+2)} \left(\sum_{k=1}^{n+1} b_k^2 + \sum_{0\leq j \leq k \leq n} a_j a_k \right).
\end{equation}
The threshold condition is now
\begin{equation}
    \delta = t^n \left|1 - i q t - p t^2\right| = t^n \sqrt{(1-p t^2)^2 + q^2 t^2 } = t^n \left[ 1 - \left(p - \frac{q^2}{2}\right) t^2 + \mc O(t^4)\right],
\end{equation}
where the $\mc O(t^3)$ terms of the amplitude only affect the modulus at $\mc O(t^4)$. This has the same form as in the previous subsection with $Q \to Q_{\mr{eff}} = p - q^2/2$. Using $\sum_{j\leq k} a_j a_k = \frac{1}{2}\big[\big(\sum_k a_k\big)^2 + \sum_k a_k^2\big]$, we find
\begin{equation}
    Q_{\mr{eff}} = \frac{1}{(n+1)(n+2)}\sum_{k=1}^{n+1} b_k^2 + \frac{\sigma_a^2}{2(n+2)}, \qquad \sigma_a^2 = \frac{1}{n+1}\sum_{k=0}^{n} a_k^2 - \left(\frac{1}{n+1}\sum_{k=0}^{n} a_k\right)^{2},
\end{equation}
and the corrected arrival time is
\begin{equation}
    \tau_{\mr{corr}} \approx \tau\left(1 + \frac{Q_{\mr{eff}}}{n}\,\tau^2\right).
\end{equation}
This shows that only fluctuations in diagonal coefficients play a role and a uniform diagonal, $a_k = a$, amounts to a global phase and leaves the modulus of overlap unchanged.

\section{Causal and non-causal models} \label{sec:sm_causality}
The causality condition of the main text imposes a restriction in the sequence of Lanczos coefficients. Certain exotic models may violate this notion of causality, such as the super-linear Lanczos coefficients obtained in the deep-Hilbert space \cite{cao_surprises2023}, or the double-scaled SYK model \cite{Bhattacharjee2023_DSSYK}.
Take as an example the family of Lanczos coefficients $\tilde b_n=\alpha\sqrt{n(n-1+\eta)}$ of the $\mathsf{SL}(2, \mbb R)$ algebra, and related SYK models, with $\eta=2\gamma/\alpha+1$ in the notation $b_n=\alpha n+\gamma+o(1)$ of~\cite{parker_universal_2019}.
For this family
\begin{equation}
    \tau_n^{2n}=\frac{n!}{(\eta)_n\,\alpha^{2n}},\qquad 
    \frac{\tau_n^{2n}}{\tau_{n-1}^{2n}}=\frac{n}{(n-1+\eta)\,\alpha^2\tau_{n-1}^2},
\end{equation}
where $(\eta)_n=\eta(\eta+1)\cdots(\eta+n-1)$ is the Pochhammer symbol. From here we find that:
if $\eta=1$ ($\gamma=0$) $\tau_n=1/\alpha$ for all $n$; causality is saturated at every site, if $\eta>1$ ($\gamma>0$) $\tau_n$ increases monotonically to $1/\alpha$ and causality holds; and lastly, if $\eta<1$ ($\gamma<0$) $\tau_n$ decreases monotonically to $1/\alpha$ and the model need not be causal.


Thus the sign of the intercept $\gamma$ in the operator growth hypothesis determines the aymptotic causality criterion.
The large-$q$ SYK model has $\eta=2/q$, $\gamma\simeq-\alpha/2$ predicting a violation of causality.
In fact, the violation already occurs at $n=2$ as its first coefficient $b_1=\mathcal J\sqrt{2/q}$ is anomalously small, so $\tau_1=\sqrt{q/2}/\mathcal J$ exceeds $\tau_2=q^{1/4}/\mathcal J$.

With the weaker condition $\tau^{\rm SL}_{n-1}\le\tau^{\rm SL}_n$, and $\tau^{\rm SL}_n=\epsilon^{1/n}\tau_n$, Eq.~\eqref{eq:causality_bound} relaxes to $b_n\le(n/\tau_{n-1})\,\epsilon^{-1/n(n-1)}$.
For SYK this removes the finite-n violation at $n=2$ whenever $\epsilon\le4/q$. However, no fixed $\epsilon$ can compensate for the asymptotic tendency generated by $\gamma<0$, so the conclusion of large-n remains unchanged.
The robust formulation is the asymptotic one
$
    \liminf_{n \to \infty} \frac{n}{b_n\tau_{n-1}}\ge1 $.
In the double-scaled limit of SYK $q\sim\sqrt N$ the Lanczos coefficients are $b_n=\alpha\sqrt{n(n-1+2/q)}$ with $\alpha=q\mathcal J\to\infty$ \cite{Bhattacharjee2023_DSSYK}. This is not a violation of linear growth, and indeed $b_n\le\alpha n$ exactly.
What diverges is the slope because $k$-locality is lost.

\section{Dynamics in full Krylov chain for short SFIM}
\label{sec:dynamics_shortSFIM}
Much of the main text dealt with examples in which the Lanczos coefficients followed a particular asymptotic behavior.
Numerically, for finite system sizes, we typically observe a growth of the Lanczos coefficients up to $n \sim \mc O(S)$ \cite{NANDY20251} where $S$ is the thermodynamic entropy of the system.
Thereafter, the Lanczos coefficients descend towards zero.
In this case, the Lanczos coefficients $b_n$ can be upper bounded by a constant $b_n \leq b^*$ so that a linear light cone $v \tau_\mr{lin}=n$ always holds asymptotically. 

\begin{figure}[ht!]
    \centering
    \includegraphics[width=.6\linewidth]{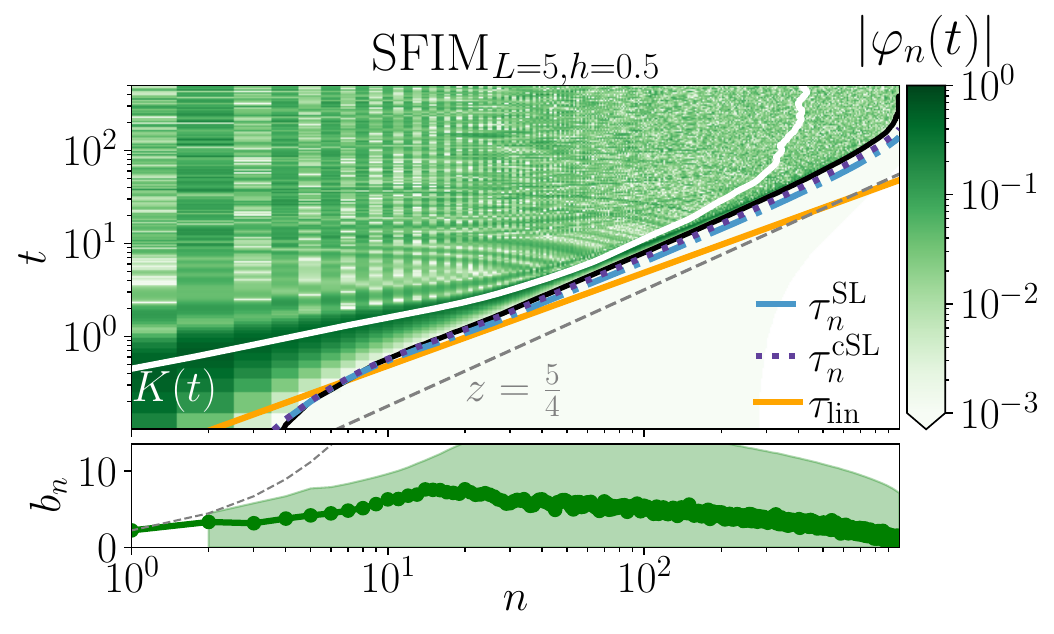}
    \caption{\textbf{Dynamics in the full Krylov chain for SFIM} with $L=5$ spins. The plot shows the same lines as in the main text: $ t_\epsilon$ (black), speed limit (blue dash-dotted), corrected SL (purple dotted), and Krylov complexity (white). Furthermore, here we show the linear light cone (orange) which holds almost always but is not tight, indeed a sub-ballistic version with $z = 5/4$ more closely reproduces the lightcone. The plot is in log-log to highlight the difference in behavior of small and large $n$ in the Lanczos, and to show clearly the power-laws. The plot also shows the Lanczos coefficients (down) (green dots), with the causality bound (green shaded area) along with the UOGH bound \eqref{eq:ogh_bound} (dashed gray line).}
    \label{fig:short_SFIM}
\end{figure}

Figure \ref{fig:short_SFIM} shows the dynamics in the entire Krylov chain for the SFIM \eqref{eq:SFIM} with few spins $L=5$ starting from the operator $\ha O_0=\hat{\sigma}^x_L$.
The full Krylov dimension is $D_K = 993$. We observe that as observed in the main text the SL performs better at short times, while the Light cone is tighter at long times.
However, note that here, since we do not use the aLC the functional dependence on the Lanczos is the same and the only difference is the prefactor.
Interestingly, SL, cSL, and LC describe well $ t_\epsilon$ with a similar power law $ t \sim n^z$ where $z\approx \frac{5}{4} >1$ (gray line), which implies that propagation is sub-ballistic due to the decrease of the Lanczos coefficients after their initial growth.
Another interesting observation is that $K(t)$ is closer to $ t_\epsilon$ in this case than in the cases described in the main text, and therefore the SL and LR provide a tighter bound for the Krylov complexity in this case.

\section{Numerical Applications}
\label{sec:sm_num_efficiency}
The bounds derived in this work are not only of fundamental importance, but they also bound the time for which a truncated Krylov space simulation is valid.
In particular, given an error tolerance $\epsilon$, each time bound dictates the maximum truncation of the dynamical generators required to keep the results within this tolerance during propagation \cite{Park1986}.
Recently added to the QuTiP library \cite{lambert_qutip_2026}, the \href{https://qutip.readthedocs.io/en/latest/guide/dynamics/dynamics-krylov.html#}{Krylov solver} can use repeated truncated Krylov subspaces with dimension $n \leq D_K$ to compute quantum dynamics, relying on the SL to bound the time at which the truncated spaces must be recalculated.
Connecting this to the asymptotic results, from the exponentially spreading light cone we see that to simulate systems with linear Lanczos $b_n \sim \alpha n$ up to times linear in system size $t\sim L$ we need an exponentially large cutoff $n \sim \exp(L)$.
Taking into account the sublinear growth of the Lanczos coefficients, a linear time in the simulation $t \sim L$ only requires a power law overhead $n \sim L^{\frac{1}{1-\delta}}$.

\subsection{Numerical Accuracy of the Bounds}
We have presented various methods to calculate a bound on information spread on the Krylov chain.
We can use these bounds in numerical applications to mark the time at which a truncated Krylov chain of dimension $n < D_K$ starts to fail in capturing the dynamics.
For the specific model of the chaotic SFIM of eq.~\eqref{eq:SFIM} with $h=0.5$, and the operator $\ha O_0 = \ha \sigma^x_L$, we compare the times obtained from the speed limit $\tau^{\mr{SL}}_n$ in eq.~\eqref{eq:speedlim}, its correction $\tau^{\mr{cSL}}_n$ in eq.~\eqref{eq:corr_speed_limit} and the Loschmidt echo method $\tau^\mr{E}_n$ described in \cite{ruffinelli_loschmidt_2022}.
To evaluate their precision, we calculate their respective differences to the numerical time $\tau^\mr{num}_n$.
It is determined by solving the whole Krylov space and comparing the evolution with that of the truncated subspace.
Then $\tau^{\mr{num}}_n$ marks (up to the size of the time step) the point at which the overlap of these two evolutions drops below $\epsilon = 1 - 10^{-10}$.
Regarding the Loschmidt method, we consider the overlap of the same operator, once evolved w.r.t. a chain of length $n-1$ and once with a length of $n$ (not the full chain).
$\tau^\mr{E}_n$ then marks the time that the overlap of these operators drops below $\varepsilon$.
\begin{figure}[ht]
    \centering
    \includegraphics[width=0.65\linewidth]{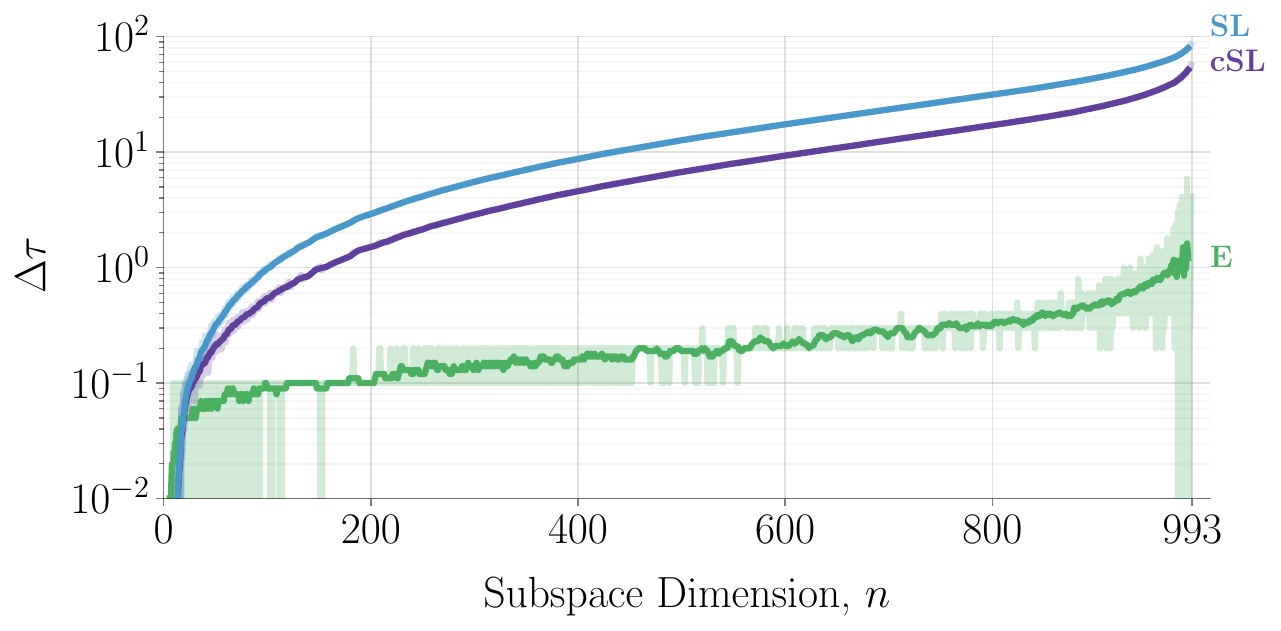}
    \caption{Accuracy $\Delta \tau$ of the bounds obtained by the speed limit (SL), its correction (cSL) and the Loschmidt echo (E), with respect to $\tau^{\mr{num}}_n$ for truncated Krylov subspaces with dimensions $n$. The maximum Krylov space dimension is $D_K = 993$. The transparent graphs show the values for each $n$, while the opaque ones show the moving average with window size 10.}
    \label{fig:accuracy}
\end{figure}

Figure \ref{fig:accuracy} shows the precision $\Delta \tau^i = \tau^{\mr{num}}_n - \tau^{i}_n$ for the various methods $i = \{ \mr{SL, cSL, E} \}$ as a function of the subspace dimension $n$.
At very low dimensions $n \lesssim 20$ all bounds are very tight since $\tau^{\mr{num}}_n$ in itself is generally small.
Moving beyond small dimensions, both SL and cSL quickly lose accuracy, showing an increasing gap as we approach the maximum Krylov space dimension $D_K = 993$.
The best accuracy is clearly presented by the Loschmidt echo method (E).
Although fluctuations make predictions difficult at low dimensions, in the intermediate regime the improvement is very clear.

\section{Derivation of the $\mc Q$-complexity bound}
\label{sm:qcomp}
Introducing the projectors over the $\mc{Q}$-eigenbasis $\mc P^\mc{Q}_q = \sum_{q_a\geq q}|q_a)(q_a|$ and the Krylov basis $\mc P^K_n = \sum_{m\geq n} |K_n)(K_n|$ it follows that $\mc P^\mc{Q}_q \mc P^K_{q/\mf c} = \mc P^K_{q/\mf c} \mc P^\mc{Q}_q= \mc P^\mc{Q}_q$ \cite{parker_universal_2019}, where $\mf c= 2 M$. This expression can be used to show that the projector over the $\mc Q$-eigenbasis is upper bounded by the projector over the Krylov basis $\mc P^\mc{Q}_q = \mc P^K_{q/\mf c} \mc P^\mc{Q}_q \mc P^K_{q/\mf c} \leq \mc P^K_{q/\mf c}$, where the inequality between superoperators follows from $\mc P^\mc{Q}_q \leq \mc I$, where $\mc I$ is the identity superoperator. The projection over one element of the $q_a$ eigenbasis can be bounded as 
\begin{equation}
    (O_t|q)(q|O_t) \leq (O_t|\mc P^\mc{Q}_q|O_t) \leq (O_t|\mc P^K_{q/\mf c}|O_t),
\end{equation}
where the first inequality follows from the fact that the terms $\sum_{q_a>q}(O_t|q_a)(q_a|O_t)\geq 0$ are positive. We now take the first non-zero term in the expansion for which $n\geq q/\mf c$, i.e., $n=\lceil q/\mf c \rceil$, to find
\begin{align} \notag 
    |(q|O_t)|^2 &\lesssim |(K_{\lceil q/\mf c \rceil}|O_t)|^2 =|\varphi_{\lceil q/\mf c\rceil }(t)
    |^2 \leq \left(\frac{t}{\tau_{\lceil q/\mf c\rceil}}\right)^{2 \lceil q/\mf c\rceil}.
\end{align}
This shows that the Krylov LR bound provides a refined bound on the $\mc Q$ complexity for which Parker et al. \cite{parker_universal_2019} proved that the average $Q$-complexity is upper bounded by the Krylov complexity $(O_t|\mc Q|O_t) = \sum_{q} q |(q|O_t)|^2 \leq K(t)$.

\section{Mapping the Krylov chain to a spin chain}
\label{sec:sm_spinchain}
The Krylov chain is a description of certain quantum dynamics as an effective one dimensional single body chain. The Krylov space with Krylov dimension $D_K$ is $\mathscr K \cong \mathbb C^{D_K}$. We consider an extension of the Krylov chain, where in every site we introduce a spin $1/2$ degree of freedom, therefore now the Hilbert space where the Krylov spin chain lives is $\mathscr{K}_\ts{sc}\cong (\mathbb C^2)^{\otimes D_K}$.  The different elements are mapped as 
\begin{subequations}
\begin{align}
	b_n |K_{n-1})(K_n|  &\mapsto b_n \ha \sigma^+_{n-1}\ha \sigma^-_n , \\
	\ha O_0 \equiv  |{K_0}) &\mapsto \ket{\uparrow_0}= \ha \sigma^+_0\ket{\downarrow \downarrow \downarrow \cdots},\\
    |K_n) &\mapsto \ket{\uparrow_n}= \ha \sigma^+_n\ket{\downarrow \downarrow \downarrow \cdots}.
\end{align}
\end{subequations} 
The flip-flop terms can be rewritten as $\ha \sigma^+_{n-1}\ha \sigma^-_n + \ha \sigma^-_{n-1}\ha \sigma^+_n = \ha \sigma^x_{n-1}\ha \sigma^x_n+ \ha \sigma^y_{n-1}\ha \sigma^y_n$, and we obtain \eqref{spinchain}.

The fundamental recurrence relation in the Krylov chain \cite{NANDY20251} encodes the action of the Liouvillian on a Krylov basis vector 
\begin{equation}
    \mbb L |K_n) = b_n |K_{n-1}) + b_{n+1}|K_{n+1}).
\end{equation}
The mapped spin chain can be seen to fulfill the same recurrence relation since $b_n \ha \sigma^+_{n-1} \ha \sigma^-_n \ket{\uparrow_n}= b_n \ket{\uparrow_{n-1}}$ while the term $b_{n+1}\ha \sigma^+_{n+1} \ha \sigma^-_n \ket{\uparrow_n} = b_{n+1} \ket{\uparrow_{n+1}}$, which gives
\begin{equation}
    \ha H\ep{\mr{sc}}\ket{\uparrow_n} = b_n \ket{\uparrow_{n-1}}+ b_{n+1}\ket{\uparrow_{n+1}}.
\end{equation}
Therefore, the action of the spin chain Hamiltonian on the single excitation sector is the same as the Krylov recurrence relation, and we can conclude that the dynamics of the single excitation of the chain is isomorphic to that of the Krylov chain since the amplitudes $\varphi_n(t)$ on every site coincide
\begin{align} \notag
    e^{i \mbb L t}|K_0) &= \sum_{n}i^n \varphi_n(t)|K_n) \\[-8pt] &\updownarrow \\[-5pt] \notag  e^{-i \ha H\ep{\mr{sc}} t}\ket{\uparrow_0} &= \sum_{n}(-i)^n \varphi_n(t)\ket{\uparrow_n}.
\end{align}
Note that our claim is that the dynamics in the two chains are the same, to go from the Krylov chain or the mapped spin chain to the original Hilbert space requires knowledge of the Krylov basis vectors $|K_n)$. 

\subsection{A looser Krylov Lieb Robinson bound}
    We follow the standard LR proof for disordered 1D models by Baldwin, Ehrenberg, Guo and Gorshkov \cite{baldwin_disorderedLR_23}, we do not consider their refinement by integrating out certain links, although it may provide an improvement of the Krylov LR bound for systems with large fluctuations in the Lanczos coefficients. Here we repeat the main steps of the basic derivation for completeness, and refer to the paper for further details \cite{baldwin_disorderedLR_23}.
    
    The framework assumes a lattice with $N$ sites in which we have a $d$-dimensional quantum system, i.e., the full Hilbert space is $\msc H = \bigotimes^{N}_{k=1} \msc H_k$ where $\mr{dim} \msc H_k=d$. We denote $\Omega$ the set of all sites of the chain and $\Lambda$ the set of all links. We consider nearest neighbors (possibly time-dependent) Hamiltonians of the form
    $
        \ha H(t) = \sum_{\ell \in \Lambda} \ha H_\ell(t),$ 
    where $\ha H_\ell$ acts on the two sites connected by link $\ell$.  This Hamiltonian may have onsite terms however they're not important since one can move to the interaction picture with respect to them. Importantly, these will be bounded by the hoppings $J_\ell$ as 
        $\lVert \ha H_\ell(t)\rVert \leq J_\ell,$ where $\lVert \bullet \rVert$ denotes the operator norm, which is an upper bound to any other norm.

One can define the basis of (orthonormal) strings
    $
        \ha X\ep{\bs \nu} = \bigotimes_{k=1}^N \ha X\ep{\nu_k}_k,$
    where $\bs \nu = (\nu_1, \cdots, \nu_N)$ with $\ha X\ep{0}_k=\Id$. We define the superoperator $\mc P_k \ha X\ep{\bs \nu} = (1-\delta_{\nu_k, 0}) \ha X\ep{\bs \nu}$ which projects on the Pauli strings which act non-trivially in site $k$. In the same way, $\mathcal P_\omega$ is the projector on any subset $\omega \subseteq \Omega$ of sites.

\begin{equation}
        \lVert \mc P_n \ha O_0(t) \rVert \leq 2 \lVert \ha O_0\rVert \left( \prod_{\ell=1}^n 4 J_\ell \right) \frac{t^n}{n!}.
\end{equation}
This means that the projection on site $n$ of the time evolution of operator $\ha O_0$ localized on site $0$ at time $t=0$, is upper bounded by the product of the hoppings connecting site $0$ to $n$. To apply the Lieb Robinson bound to our Krylov spin chain we want to bound the $n$-th site projection of the time evolution of a certain operator  $\ha O^{(n)}_t = (\ha K_n, \ha O_t)$. Assuming that the $\ha K_n$ operator is traceless $\ha \Tr(\ha K_n)=0$, its norm is bounded by unity $(K_n, K_n)\leq 1$, and that it only acts nontrivially on the $n$-th site, we can simplify $\ha O^{(n)}_t = (\ha K_n,\mc P_n \ha O_t) $. In the spin chain we choose the operator $\ha K_n = \ha \sigma^+_n$ since it fulfills: $\Tr(\ha \sigma^+_n)=0, \, (\sigma^+_n, \sigma^+_n)=\frac{1}{D_\mr{sc}}\Tr(\ha \sigma^-_n\ha \sigma^+_n)\leq 1$. The projection can be bounded through Cauchy-Schwarz $|O^{(n)}_t|^2 \leq (\ha K_n, \ha K_n)(\mc P_n \ha O_t, \mc P_n \ha O_t) \leq \lVert \mc P_n \ha O_t\rVert^2$. Finally, in the mapped spin chain the relevant upper bound on the interaction terms is given by the Lanczos coefficients $\lVert \ha H_\ell\rVert\leq b_\ell$, and we find the bound
\begin{equation}
	|O^{(n)}_t|\leq  \lVert \mc P_n \ha O_t\rVert \leq 2 \lVert \ha O_0\rVert b_1 \cdots b_n \frac{(4 t)^n}{n!}.
\end{equation}

\end{document}